
\input phyzzx

\nopagenumbers
\line{\hfil CU-TP-592}
\vglue .4in
\centerline {\twelvebf Radiation from a Moving Scalar Source}
\vskip .3in
\centerline{\it Hai Ren and Erick J. Weinberg }
\vskip .1in
\centerline{Physics Department}
\centerline{Columbia University }
\centerline{New York, New York 10027}
\vskip .4in
\baselineskip=20pt
\overfullrule=0pt
\centerline {\bf Abstract}
\medskip

	We study classical radiation and quantum bremsstrahlung
effect of a moving point scalar source. Our classical analysis
provides another example of resolving a well-known apparent
paradox, that of whether a constantly accelerating source radiates
or not. Quantum mechanically, we show that for a scalar source
with arbitrary motion, the tree level emission rate of scalar
particles in the inertial frame equals the sum of emission and
absorption rates of zero-energy Rindler particles in the Rindler
frame. We then explicitly verify this result for a source
undergoing constant proper acceleration.

\vskip 1.1in
\noindent\footnote{}{\twelvepoint This work was supported in part by
the US Department of Energy }

\vfill
\eject

\baselineskip=20pt
\pagenumbers
\pageno=1

\def\pr#1#2#3#4{Phys. Rev. D{\bf #1}, #2 (19#3#4)}

\def\refmark#1{[#1]}
\def\sqr#1#2{{\vcenter{\hrule height.#2pt
        \hbox{\vrule width.#2pt height#1pt \kern#1pt
                \vrule width.#2pt}
        \hrule height.#2pt}}}
\def\square{\mathchoice\sqr64\sqr64\sqr{2.1}3\sqr{1.5}3}
\def\iint{{\int^{+\infty}_{-\infty}}}
\def\p{\partial}
\def\gd{\delta}

\def\w{\omega}

\def\bv{{\bf v}}
\def\bk{{\bf k}}
\def\bx{{\bf x}}
\def\kper{{k_\bot}}
\def\tg{\theta}
\def\gl{\lambda}
\def\tag{\p_\alpha\phi\p^\alpha\phi}

\chapter{ Introduction}

     The problem of a uniformly accelerating electric charge gives
rise to a well known apparent paradox.  Since the charge is
accelerating, it should radiate.  However, by the principal of
equivalence, the situation should be equivalent to that of a static
charge in a uniform gravitational field, which certainly does not
radiate.  The resolution lies in the recognition that only a portion
of Minkowski space-time is accessible to a uniformly accelerating
observer comoving with the charge.  A careful analysis at either the
classical \Ref\Boulware{D.G.~Boulware, Ann.Phys.(N.Y.) {\bf 124},
169 (1980).} or quantum \Ref\Higuchi{A.~Higuchi, G.E.A.~Matsas, and D.~
Sudarsky. \pr{45}{R3308}92; \pr{46}{3450}92.} level then shows that it
is possible for this coaccelerating observer to conclude that there is
no radiation, even though a static observer sees the charge radiate.

      In this paper, we consider a closely related problem, that of a
uniformly accelerating source coupled to a massless scalar field $\phi$
through the Lagrangian
$$ {\cal L} = -{1\over 2} \partial_\mu \phi \partial^\mu \phi + \rho\phi
   \eqn\i $$
As in the electromagnetic case, a static observer would expect the
source to radiate, while a coaccelerating observer would not.  Although
there are number of differences from the electromagnetic
cases, including the detailed form of the radiation, we show that for this
case also the views of the two observers can be reconciled.

      In Minkowski coordinates $(t,x,y,z)$ the source
is uniformly accelerating, following the trajectory
$$ x_s^\mu(s) = (a^{-1}\sinh as, 0, 0, a^{-1}\cosh as)
   \eqn\ii $$
where $s$ is the proper time of the source.
For describing the observations of the coaccelerating observer, it is
convenient to use Rindler\Ref\Rindler{W.~Rindler, Am. J. Phys. {\bf
34} 1174 (1966).} coordinates $(\tau, x, y, \xi)$ defined by
$$   t = {e^{a\xi}\over a}\sinh a\tau \   , \qquad
 z = {e^{a\xi}\over a}\cosh a\tau
   \eqn\iii $$
In terms of these flat Minkowski metric takes the form
$$   ds^2 = e^{2a\xi}(- d\tau^2 + d\xi^2) + dx^2 + dy^2
   \eqn\iv $$
In the Rindler coordinates the source is stationary at
the point $x=y=\xi=0$.

    The Rindler coordinates cover only the wedge $z >|t|$ (region~I) of
Minkowski space-time, which is the only part completely accessible to an
observer comoving with the source.  Region~II ($t > |z|$) is always outside
the observer's past light cone; thus, although he can send signals to this
region, he can never observe events there.   Similarly, he can receive
signals from region~III ($t< -|z|$), but can never send signals there.
Finally, he can have no communication with region~IV (($z< -|t|$).

     Since the source of a massless scalar field need not be conserved,
its time-dependence must be specified.  We take it to have constant
magnitude in its rest frame:
$$   \rho = q  \delta(x) \delta(y) \delta(\xi)
   \eqn\sourcerind$$
Lorentz contraction of the volume then makes the magnitude of the source
time-dependent in the inertial frame:
$$ \eqalign{ \rho &=   {q \over a\sqrt{t^2 + a^{-2}} }\,
     \delta(x)\,\delta(y)\, \delta(z - \sqrt{t^2 + a^{-2}}) \cr
      &=  q\sqrt{1 - \bv_s^2}\,\gd^{(3)}\,(\bx - \bx_s(t))}
   \eqn\vi $$

        In Sec.~II we consider the problem from a classical point of
view.  For the electromagnetic case, several features explain the
failure of the comoving observer to observe radiation.  It turns out
that within region~I it is never possible to clearly distinguish a
radiation field distinct from the expected Coulomb field.  One might
instead examine the flow of energy, but this turns out to be entirely
into regions inaccessible to the comoving observer.  We find that
these features are reproduced in the scalar case.  A difference which
we find is that, although the retarded potential method fails to give
the correct result for the electromagnetic case \Ref\Bondi{H.Bondi and
T.Gold, Proc. Roy. Soc. A {\bf 299}, 416 (1955).}, it does give a true
solution of the field equations in the scalar case.  We examine this
in some detail.

    In the quantum theory the radiation due to the accelerated charge
appears as the emission of quanta by a bremsstrahlung effect.  Thus,
one might ask whether the static and the comoving observers both see
emission of particles.  However, this is not quite the right question.
It is well known that an observer who is static in Rindler coordinates
has a different definition of particle than does a static Minkowski
observer, and interprets the Minkowski vacuum as a
Fulling-Davies-Unruh (FDU) thermal bath of
many-particle states \REFS\Fulling{S.A.Fulling, \pr7{2850}73.}
\REFSCON\Davies{P.C.W.Davies, J.~Phys.~A{\bf 8}, 609 (1975)}
\REFSCON\Unruh{W.G.Unruh, \pr{14}{870}76.}\refsend.  The underlying
reason for this difference is
that modes which have positive frequency with respect to Minkowski
time are linear combinations of Rindler modes with both positive and
negative frequencies.  A consequence of this fact is that what a
Minkowski observer calls emission of a quantum can appear to a Rindler
observer as either emission or absorption of a quantum.  In Sec.~III
we first show that for any source confined to region~I
the Minkowski emission rate
is equal to the sum of an emission and an absorption rate
calculated by the Rindler observer.   We then verify by explicit
calculation that the Minkowski bremsstrahlung rate due to a uniformly
accelerated source is precisely equal to the the sum of the rates
for emission and absorption of zero-energy quanta by a static source
in the thermal bath.  This calculation parallels
that of Ref.~\Higuchi\ for the electromagnetic case.

	In the appendix, we study some general properties of classical
radiation from a moving point scalar source.

\chapter{Classical Radiation from a Uniformly Accelerated Scalar Source }

    In this section we consider the problem from a classical point of
view.  Our treatment parallels that of Boulware \refmark{\Boulware} for the
electromagnetic case.

   The first step is to determine the classical field generated by our
source.  We do this in Minkowski coordinates, with the source given by
Eq.~\iv.  Solving the wave equation $\square \phi =-\rho$ by the
retarded potential method gives
$$ \phi(x) = {q\over 2\pi}\iint ds\theta(t - x^0_s(s))\gd((x -
x_s(s))^2)
		\eqn\i   $$
where $x_s^\mu(s)$ is given by Eq.~\ii. Thus
$$ \phi(x) = {q\over 4\pi R}	\theta(t + z)
		\eqn\ii    $$
where
$$ R = {a\over 2}\left[ (X^2 -a^{-2})^2 + 4a^{-2} \rho^2
\right]^{1/2}
   \eqn\iii $$
with $\rho^2 = x^2 + y^2 $ and $X^2 = x^\mu x_\mu = \rho^2 +z^2 -t^2$.
It is easy to verify that this is a solution of the field equation, even
on the plane $t+z=0$ on which its derivatives are singular.

     In the electromagnetic analogue, the fields obtained from the
Li\'enard-Wiechert potentials do not satisfy Maxwell's equations
along the surface $t+z=0$, but instead differ from the actual
solutions by  terms  proportional to $\delta(t+z)$.  These terms can
be motivated by a limiting process suggested by
Bondi and Gold \refmark{\Bondi}.
Consider an electric charge which is at rest at $z=1/a$ until
$t=0$, and is uniformly accelerated after that.  Now apply
the Lorentz transformation
$$ z \longrightarrow z\cosh\alpha + t\sinh\alpha , \qquad
t \longrightarrow t\cosh\alpha + z\sinh\alpha
		\eqn\iv    $$
thus going to a frame in which the charge has a constant
negative initial velocity and in which the uniform acceleration
begins at $t= -a^{-1} \sinh \alpha$.  In the limit $\alpha \rightarrow
\infty$, the initial velocity of the charge approaches the speed
of light, and the time at which the uniform acceleration begins
goes to $-\infty$.  In this limit the Coulomb field of the
initially static charge is Lorentz tranformed into a delta
function field along the surface $t+z=0$.

      Even though the retarded potential method gives a solution
for the scalar case, one might wonder if this limiting
procedure might lead to an additional contribution.  This is readily
examined.   A  point source which is at rest at $z=1/a$ until
$t=0$ and uniformly accelerated thereafter gives rise to a field
$$ \bar\phi(x) = {q\over 4\pi} \left[ {1\over r}\theta (r-t)
   + {1\over R}\theta (t-r) \right]
     \eqn\v $$
where $r = [\rho^2 + (z- a^{-1})^2 ]^{1/2}$.  (Note that $t=r$ implies
that $R=r$, so that this solution is continuous everywhere.)
Applying the transformation
\iv\ gives
$$ \bar\phi(x) = {q\over 4\pi} \left[ {1\over r'}\theta (\lambda -t-z)
   + {1\over R}\theta (t+z -\lambda) \right]
     \eqn\vi $$
where
$$ \lambda = e^{-\alpha} \left[  a^{-1}
    + {\rho^2 \over a^{-1} -(z-t)e^{-\alpha} } \right]
     \eqn\vii $$
$$ r' = \left[ \rho^2 + (z \cosh \alpha + t\sinh \alpha - a^{-1} )^2
\right]^{1/2}
     \eqn\viii $$
and $R$ is as before.  Since $\lambda$ vanishes in the
limit $\alpha \rightarrow \infty$, while $ r' \sim |z + t| e^\alpha$
for $z+t < 0$, we see that $\bar\phi(x)$
approaches the field \ii\ of the uniformly accelerating source for any
point with $t+z \ne 0$. The behavior on the null plane $t+z =0$ is
somewhat curious.  On this plane, $r' = (\rho^2 + a^{-2})^{1/2} +
O(e^{-\alpha})$, while $R = (a/2)(\rho^2 + a^{-2})$.  Consequently,
the limits $t \rightarrow -z$ and $\alpha \rightarrow \infty$ do not
commute.  In fact, the
field is rapidly varying in a region of width of order $e^{-\alpha}$
about this plane, with
$$ {\bar\phi(z+t=\lambda) \over \bar\phi(z+t=0)}
   = {2 \over \sqrt{1 +a^2 \rho^2}}
     \eqn\ix $$

    In the electromagnetic case, this limiting process leads
to a delta function contribution to the field strengths which is not
obtained by the Li\'enard-Wiechart method and which is needed to
satisfy Maxwell's equations on the plane.  In the scalar case, the
fields obtained by the two methods differ at most by a finite amount
on the $t+z=0$ plane. The differences seem a bit more
significant when we consider the energy-momentum tensor $ T^{\mu\nu}
= \p^\mu\phi\p^\nu\phi - {1\over 2}\eta^{\mu\nu}
\p_\alpha\phi\p^\alpha\phi $.  With the retarded potential solution
\ii , the step functions give rise to singular contributions
(proportional to squares of delta functions) to $T^{\mu\nu}$.  While
the limiting method gives a way of defining these singular
contributions more precisely, it is not at all clear that in the limit
$\alpha\rightarrow\infty$ they agree with any reasonable definition of
the former case.  This is not surprising. Within the region $0\le t+z
< \lambda \sim e^{-\alpha}$ (and thus on the null plane, in the limit)
it is possible to distinguish a source which was initially moving with
constant velocity from one which has always been uniformly
accelerating.  The essential point is that the comoving observer's
analysis of the situation depends only on the values of the field
within region~I, but not on its boundary, so that the ambiguity in
defining $T^{\mu\nu}$ on the boundary is immaterial to the problem of
reconciling the views of the static and the comoving observers.

      We can also examine the effects of this limiting procedure
on the gradients of the field $\p^\mu\phi(x)$, which constitute
the energy-momentum
tensor. One can do this by applying the limiting procedure directly
to $\p^\mu\phi$, or alternatively, by simply differentiating
Eq.~\vi
$$ \p^\mu\bar\phi(x) = {q\over 4\pi} \left[
\theta (\lambda -t-z) \p^\mu\left({1\over r'}\right)
   + \theta (t+z -\lambda) \p^\mu\left({1\over R}\right) \right]
     \eqn\x $$
In the limit $\alpha \rightarrow \infty$, we get a $\delta$-function
contribution from the Lorentz-transformed Coulomb field. These
$\delta$-function terms are the same as those one would
get by directly differentiating the retarded potential solution \ii .

      We now explicitly calculate the energy-momentum tensor for the
case of the uniformly accelerating source.  Since we will need it only
for $z+t >0$, we ignore the singular contributions.  In this region a
straightforward calculation in Minkowski coordinates gives
$$ \eqalign {T^{\mu\nu} &= {q^2a^2 \over 16\pi^2} \left\{
  {1\over R^4} \left(x^\mu x^\nu -{1\over 2} \eta^{\mu\nu}X^2 \right)
     \right.\cr
  &\qquad \left. + {1\over R^6} \left[ a^{-2} \rho^\mu\rho^\nu
    +{1\over 2} (X^2 -a^{-2})(x^\mu\rho^\nu + \rho^\mu x^\nu)
    - \rho^2 x^\mu x^\nu \right] \right\} }
   \eqn\xxx $$
where $\rho^\mu \equiv (0,x,y,0)$ and the $\theta(t + z)$ factors have
been suppressed.

      It is also straightforward to obtain the components of
$T^{\mu\nu}$ with respect to the Rindler coordinates.  In particular,
the components $T^{\tau j}$, which correspond to the energy flux seen
by a comoving observer, vanish everywhere.  This follows from the fact
that the field \ii\ is static when written in terms of Rindler
coordinates.  Alternatively, one can simply transform from the
Minkowski result, using the formulas
$$  T_{\tau j} = {\p x^\mu\over \p \tau} T_{\mu j} =0
    \eqn\xii $$
for $j=1$ or 2 and
$$ T_{\tau \xi} = {\p x^\mu\over \p \tau}{\p x^\nu\over \p \xi}T_{\mu\nu}
     =0
    \eqn\xiii $$
with indices $\mu$ and $\nu$ referring to Minkowski components.

    While the comoving observer sees no flow of energy, and thus
no radiation, matters are not so simple for the static observer.
Thus, let us calculate the power radiated by the source, as seen in
Minkowski coordinates.  This quantity is Lorentz invariant, and is
most easily calculated in a frame where the source is instantaneously
at rest.   Furthermore, because the acceleration of the source is
uniform, the radiation should be the same at all points along the
world-line of the source.  We therefore look along the forward
light-cone of the point $x=y=t=0, z=a^{-1}$ at which the source is at
rest.  On this light-cone, the components of
the energy flux are
$$ T^{tj} = {q^2 \over 16\pi^2} \left[{a^2 \cos^2 \theta \over r^2}
     + {a \cos \theta \over r^3} \right] \hat{\bf r}^j
     \eqn\xvi $$
where $\hat{\bf r}$ is a unit three-vector from the point $x=y=0,
z=a^{-1}$ to the field point, $r$ is the three-dimensional
distance between these two points, and $\theta$ is the angle between $\hat
{\bf r}$ and the $z$-axis.  By integrating over a sphere of radius $r$,
one finds that the energy flux along this light cone at time $t=r$ is
$$ \eqalign {\int dS_j T^{tj} &= {q^2\over 16\pi^2}\int d\Omega
  \left[ a^2\cos^2\theta + {a \cos \theta \over r}\right] \cr
     &= {q^2a^2\over 12\pi} }
		\eqn\xvii  $$
We show in the Appendix that Eq.~\xvii\ is exactly the result
expected for the power radiated by an uniformly accelerating source.

     Although this result is suggestive of radiation, the real test is
whether the energy in the field changes over time.  This can be
addressed by calculating the net energy flux through a closed
three-dimensional hypersurface. In particular, let us consider the
region \refmark{\Boulware} given by $ z > |t|+
\epsilon$ (with the limit $\epsilon\rightarrow 0$ understood) and
$z_1 < z < z_2$, with $z_1 < a^{-1} < z_2$, and calculate the net
energy flux out of this region, as seen by a static observer using
Minkowski coordinates.
Because $T^{tz}$ is an odd function of $t$, the total flux through either
of the surfaces $z=z_1$ or $z=z_2$ vanishes.  The flux
through the surfaces $ z = \pm t + \epsilon$ is
$$\eqalign{\int d^2\rho \int^{z_2}_{z_1}dz(T^{tt} \mp T^{tz})
  (t=\pm z,x,y,z)
&= {q^2a^6\over 2\pi^2}(z_2 - z_1)\int d^2\rho {\rho^2\over
(1 + a^2\rho^2)^4}\cr  &= {q^2a^2 \over 12\pi} (z_2 - z_1)}
		\eqn\xviii     $$
Thus, the energy flowing in through the surface $z = -t + \epsilon$ is
exactly equal to that flowing out through the surface $z = t +
\epsilon$.  Hence, the static observer, like the comoving observer,
will conclude that there is no net energy production in the region.

\def\V{{\cal V}}
     In fact, the total radiation flowing out of any closed
three-dimensional hypersurface symmetric in $t$ and confined to region
I is zero.
To see this, let $\V$ be a four-dimensional spacetime volume
with three-dimensional boundary $\p \V$, whose outward normal
we denote as $n_\mu$. The total flux
through $\p \V$ is
$$\eqalign{\oint_{\p \V}d^3x&T^{t\mu}n_\mu = \int_\V d^4x\p_\mu
T^{t\mu} = \int_\V d^4x(\p^t\phi(x))(\square\phi(x))   \cr
&= -q\int_\V d^4x\sqrt{1 - \bv_s^2}\,\delta^{(3)} (\bx - \bx_s(t))
     \p^t\phi(x)   \cr
&= -q\int dt \sqrt{1 - \bv_s^2}\,\p^t\phi(x)\Big |_{\bx=\bx_s(t)}
= 0   }
		\eqn\xix   $$
since $\sqrt{1 - \bv_s^2}$ is even in $t$ while $\p^t\phi$ is odd
in $t$.

     As an alternative to looking for energy flow as evidence of
radiation, one might also examine the field (or, more precisely, its
gradient) to see whether it is possible to distinguish separate
Coulomb and radiation components.  Along the forward
light-cone of the source in its instantaneous rest frame,
the spatial gradients of the field are
$$  \partial_j \phi = {q \over 2\pi}
  \left[ {1\over r^2} + { a\cos\theta \over r} \right] \hat{\bf r}^j
    \eqn\xx $$
where the notation is the same as in Eqs.~\xvi\ and \xvii .  Using their
$r$-dependence to identify the two terms in brackets as Coulomb and
radiation components, respectively, we see that the latter
dominates when $|ar\cos \theta| \gg 1 $.  However, in region~I the
condition $z > |t|$ implies that
$$ {\rm radiation~ field \over Coulomb~field} =
     ar \cos\theta < {\cos\theta \over 1 -\cos\theta}
   \eqn\xxi $$
Hence, in the region accessible to the comoving observer the radiation
field can be dominant only for $\theta$ very close to zero.  Even in
that small region the issue is confused.  Let $l$ be the distance from
a given point in region~I to the world line of the source,  measured to
the point on the world-line where the spacelike separation is
greatest.  For $\cos\theta$ near unity one finds that $l^2 \approx
2r/a$, so that the radiation component in Eq.~\xx\ takes on the Coulomb
form, but with $l$ taking the place of $r$.


\chapter{ Scalar Bremsstrahlung and the FDU Thermal Bath}

    In the quantum theory radiation is not continuous, but rather is
a series of discrete events --- the emission of discrete quanta ---
each corresponding to a change in the state of the quantum field.  The
static and the uniformly accelerating observers do not agree on the
initial state of the field --- the Rindler observer interprets the
Minkowski vacuum as a thermal bath of many-particle states --- but
they should agree on whether the state changes.   To show how this
works, we first show that for an arbitrary source confined to region
I, the emission rate seen by the static observer is equal to the sum
of the emission and absorption rates measured by the accelerating
observer.  We then specialize to the uniformly accelerating source of
Eq.~\sourcerind, and verify this result by explicit calculations.

    We begin by expanding the quantum field in terms of normal modes.
To do this, we need a set of solutions of the scalar wave equation
$\square\phi=0$ which form a complete set on a spacelike slice through
space-time.  A convenient choice for such a slice is the hypersurface
given by $t=0$ in Minkowski coordinates.  This slice lies partly in
region~I (where it is specified by $\tau=0$) and partly in region~IV.
For Rindler coordinates, a full set of modes comprises a
complete set in the Rindler coordinates for region~I, supplemented by a
similar complete set in terms of the analogous coordinates for
region~IV.

    Consider first the decomposition appropriate to Rindler space.
Let
$$ \phi(x) = \phi_R(x) + \phi_L(x)
    \eqno\eq $$
where $\phi_L$ ($\phi_R$) vanishes if $x$ lies in region~I (region~IV).
Because the Rindler metric is independent of $x$, $y$, and $\tau$,
the modes in region~I can be chosen to be of the form
$$  f_{k_x,k_y,\omega}  = e^{ik_x x + ik_y y -i \omega \tau}
    h_{k_x,k_y,\omega}(\xi)
     \eqn\rindmodes $$
The field in region~I can then be expanded as
$$ \phi_R(x)  = \iint dk_x\iint
	dk_y\int^{+\infty}_0 d\omega\left[a_{k_x,k_y,\omega}
f_{k_x,k_y,\omega}(x) + h.c.\right]
     \eqno\eq$$
where, as usual, the positive and negative frequency modes have been
separated.   A similar decomposition for $\phi_L(x)$ can be made in
region~IV.

    The normalization of the $f_{k_x,k_y,\omega}(\xi)$ can be fixed by
requiring that
$$ (f_{k_x,k_y,\omega},f_{k_x',k_y',\omega'})_{Rind} = F(\omega)
      \delta(\w - \w')\delta(k_x -k_x')\delta(k_y - k_y')
    \eqn\rindnorm $$
where for any two functions $f(x)$ and $g(x)$ we define
$$ (f,g)_{Rind} \equiv
   i \int d^3x {\sqrt h} n^\mu
       \left[ f^*(x){\mathop{\partial_\mu}^\leftrightarrow}g(x) \right]
       \eqno\eq $$
Here the integration is over the region~I hypersurface $\tau=0$, with
$n^\mu = (e^{-a\xi}, 0, 0, 0) $ being the normal to that hypersurface
and $h_{\mu\nu} = g_{\mu\nu} + n_\mu n_\nu $ the induced
three-dimensional metric.  If $f$ and $g$ are solutions of the scalar field
equation, then $(f,g)$ is independent of the choice of the spacelike
hypersurface.  The choice for the function $F(\omega)$ determines the
commutation relations of the creation and annihilation operators $a$
and $a^\dagger$ and, through these, the density of one-particle states;
physical results are insensitive to the particular choice made.

     In Minkowski coordinates the field is usually expanded in plane
waves.  For the present purposes it is more convenient to choose the
modes to have definite transverse momenta $k_x$ and $k_y$, but not
definite $k_z$ or frequency.  Although the modes need not each have a
single frequency, they should be chosen so that their Fourier
components (with respect to the Minkowski time $t$) are either all
positive frequency or all negative frequency.  This ensures that the
associated operators in the mode expansion of the field have
simple interpretations as particle creation and annihilation
operators.  Specifically, we choose for the Minkowski modes the
linear combinations of Rindler modes
$$  g^{(1)}_{k_x,k_y,\omega}(x)
     = A(\omega) \left[ f_{k_x,k_y,\omega}^{(R)}(x)
    + e^{-\pi\omega/a} \left( f_{-k_x,-k_y,\omega}^{(L)}(x)\right)^* \right]
    \eqn\minkmodea $$
and
$$  g^{(2)}_{k_x,k_y,\omega}(x)
     = B(\omega) \left[ f_{k_x,k_y,\omega}^{(L)}(x)
    + e^{-\pi\omega/a} \left( f_{-k_x,-k_y,\omega}^{(R)}(x)\right)^* \right]
    \eqn\minkmodeb $$
which were shown by Unruh \refmark{\Unruh} to have only positive
frequency Fourier
components;  here superscripts $R$ and $L$ refer to the modes defined
in region~I and region~IV, respectively.

   The normalization of these can be fixed by requiring
$$ ( g^{(i)}_{k_x,k_y,\omega}, g^{(j)}_{k_x',k_y',\omega'},f)_{Mink}
      = F(\omega)
   \delta_{ij}   \delta(\w - \w')\delta(k_x -k_x')\delta(k_y - k_y')
    \eqn\minknorm $$
where
$$ (f,g)_{Mink} \equiv
   i \int d^3x f^*(x){\mathop{\partial_t}^\leftrightarrow}g(x)
       \eqno\eq $$
with the integration is over the hypersurface $t=0$.  (Again, for
solutions of the field equation, $(f,g)$ does not depend on the choice
of the constant time surface.)  Note that if the same choice of
$F(\omega)$ is made in Eqs.~\rindnorm\ and \minknorm, then for a pairs
of functions
with support only in region~I (or only in region~IV) $(f,g)_{Mink} =
(f,g)_{Rind}$, while if $f$ has support in region~I and $g$ in
region~IV, $(f,g)_{Mink} =0$.  Note also that $(f^*, g^*) = - (f,g)$.
It follows that if the Rindler modes \rindmodes\  are
properly normalized, then the Minkowski modes \minkmodea\ and
\minkmodeb\  will be normalized if
$$ A(\omega) = B(\omega) = \left[ 2\sinh(\pi \omega/a) \right]^{-1/2}
     \eqno\eq $$

      To lowest order in perturbation theory, the amplitude that a
source $\rho(x)$ leads to the creation from the vacuum of a quantum in
the state $|n\rangle$ is
$$  {\cal A} =
    \langle n| i\int d^4x\sqrt{g} \rho(x) \phi(x) |vac \rangle
       \eqno\eq $$
The total emission probability is obtained by squaring the amplitudes
and summing over all one-particle final states.  After expanding the
fields in terms of the modes \minkmodea\ and \minkmodeb,
one finds that the probability for emission of a
Minkowski quantum with transverse momentum $(k_x, k_y)$, assuming that the
field was initially in the Minkowski vacuum, is
$$  dP_{k_x,k_y}^{Mink} = \sum_j \int_0^\infty d\omega F^{-1}(\omega)
   \left| \int d^4x \sqrt{g} \rho(x) g^{(j)}_{k_x,k_y,\omega}(x) \right|^2
      \eqno\eq $$
Now let us assume that $\rho(x)$ vanishes outside of region~I.  The
contribution from the $g^{(1)}$ modes is then
$$ \eqalign{  dP_{k_x,k_y}^{Mink,1}
    &=  \int_0^\infty d\omega F^{-1}(\omega)|A(\omega)|^2
   \left| \int d^4x \sqrt{g} \rho(x)  f_{k_x,k_y,\omega}^{(R)}(x)
                \right|^2 \cr
     & = \int_0^\infty d\omega \left [{1\over e^{2\pi\w/a} - 1} + 1\right ]
        {dp_{k_x,k_y}^{Rind}(\omega)\over d\omega}   }
      \eqn\Mproba $$
where $dp_{k_x,k_y}^{Rind}/d\omega$ is the emission
probability per unit frequency range in the Rindler
vacuum.  The factor multiplying $dp_{k_x,k_y}^{Rind}/d\omega$ converts this
vacuum emission
probability to the sum of the spontaneous and the induced emission
probabilities in a thermal state with temperature $a/2\pi$, which is how
the Rindler observer interprets the Minkowski vacuum,   Similarly, the
contribution from the $g^{(2)}$ modes is
$$ \eqalign{  dP_{k_x,k_y}^{Mink,2} &=  \int_0^\infty d\omega
       F^{-1}(\omega)|B(\omega)|^2   e^{-2\pi\omega/a}
   \left| \int d^4x \sqrt{g} \rho(x) (f_{-k_x,-k_y,\omega}^{(R)}(x))^*
          \right|^2 \cr
     &= \int_0^\infty d\omega \left[{1\over e^{2\pi\w/a} - 1}\right]
      {dp_{-k_x,-k_y}^{Rind}(\omega)\over d\omega}}
      \eqn\Mprobb$$
which is the probability for absorption of a quantum with transverse
momentum $(-k_x, -k_y)$ in the same thermal state.  Adding
Eqs.~\Mproba\ and \Mprobb, we obtain the desired result.

     Let us now verify this result for the special case of the
constantly accelerating source.  The total transition
probability is infinite, since the source is present for all
times.  We therefore calculate the emission and absorption rates,
defined as transition probabilities per unit proper time of the
source.

     We start with the Minkowski calculation.  Instead of the modes
\minkmodea\ and \minkmodeb, we work with plane wave modes.  A standard
calculation then gives the emission rate
\def\dkz{{dk_z \over (2k_0)}}

$$ \eqalign {  dW^{Mink}_{k_x,k_y}
     &=  {1\over (2\pi)^3 T} \iint\dkz
	\left|\int d^4 x \rho(x)e^{i(k_0 t - \bk\cdot\bx)}\right|^2   \cr
       &={q^2\over (2\pi)^3 T}
       \iint\dkz\iint d\tau '\iint d\tau ''   \cr
	&\qquad\qquad \times \exp\left [
	-i{k_z\over a}(\cosh a\tau ' - \cosh a\tau '') + i{k_0\over a}
	(\sinh a\tau ' - \sinh a\tau '')\right ]    }
		\eqno\eq     $$
where $k_0 \equiv |{\bf k}| \equiv (k_z^2 + \kper^2)^{1/2}$ and $T$
represents the (infinite) total proper
time along the trajectory of the source.   If we write
$\tau \equiv (\tau' + \tau'')/2 $ and $\sigma \equiv (\tau'- \tau'')/2 $
and define
$$  \eta = \cosh^{-1}\left[ {k_0\over\kper} \cosh a\tau
    - {k_z\over\kper} \sinh a\tau \right]
		\eqno\eq  $$
this expression can be rewritten as \Ref\integrals{I.S.~Gradshteyn and
I.M.~Ryzhik, {\it Table of Integrals, Series, and Products} (Academic,
New York, 1980) p.~959;
W.~Magnus, F.~Oberhettinger, and R.P.~Soni, {\it Formulas and Theorems
for the Special Functions of Mathematical Physics} (Springer, New York
1966) p.~96.}
$$ \eqalign {dW^{Mink}_{k_x,k_y} &= {q^2 \over (2\pi)^3} {1\over T}
   \iint d\tau   \iint d\eta \iint d\sigma
\exp\left [{2i\kper \cosh\eta \over a}\sinh a\sigma\right ] \cr
     &= {q^2 \over 4\pi^3a} \iint d\eta
         K_0 \left({2\kper \cosh\eta \over a} \right) \cr
    &={q^2\over 4\pi^3a} \left |K_0\left({\kper\over a}\right)\right |^2   }
		\eqn\minkemit    $$
Here we have cancelled the factor of $T$ by the integral $\iint
d\tau$.

    We want to compare this result with the Rindler emission and
absorption rates.  There is a problem here because the source is
static in Rindler coordinates.  Normally, one would then conclude that
there was also no induced emission.  However, because the density of
quanta in the FDU thermal bath diverges as the frequency goes to zero,
matters are more subtle.  We adopt the approach of Ref.~\Higuchi\ and replace
the static source of Eq.~\sourcerind\  by the time-dependent source
$$ \rho = {\sqrt 2}q\cos E\tau\delta(\xi)\delta(x)\delta(y)
		\eqn\oscsource    $$
The limit $E\rightarrow 0$ will be taken at the end of the
calculation.

     To proceed further we need an explicit expression for the Rindler
modes.  If the choice $F(\omega)=1$ is made in Eq.~\rindnorm, then the
appropriately normalized modes are \refmark{\Fulling}
$$ f_{k_x,k_y,\omega}^{(R)} =  {1\over 2\pi^2}
     \left [{\sinh(\pi\w/a)\over a}\right]^{1\over 2}
	K_{i{\w\over a}}\left ({k_\bot\over a}e^{a\xi}\right)e^{ik_xx +
	ik_yy -i\w\tau}
		\eqno\eq   $$
where $K_\nu(z)$ is the Bessel function of imaginary argument.  Using
this expression together with Eq.~\oscsource\ for the source,
and comparing with Eq.~\Mproba, we see that the emission
rate per unit frequency range in the Rindler vacuum is
$$   {dw_{k_x,k_y}^{Rind}(\omega)\over d\omega} =
        {q^2\over 2\pi^2 a}{1\over T}\sinh(\pi\w/a)
      \left[K_{i{\w\over a}}\left ({k_\bot\over a}\right)\right]^2
       \left|\iint d\tau e^{-i\omega\tau} \cos E\tau\right|^2
        \eqno\eq $$
where $T$ represents the length of the total time interval.  The
integrals over $\tau$ each give factors of $\pi \delta(E - \omega)$.
(There are also terms involving $ \delta(E + \omega)$; we omit these,
since $E$ and $\omega$ are both positive.)  Writing $2\pi \delta(0) =
T$ allows us to cancel the factor of $1/T$, and leaves us with
$$   {dw_{k_x,k_y}^{Rind}(\omega)\over d\omega} =
        {q^2\over 4\pi^2 a}\sinh(\pi\w/a)
      \left[K_{i{\w\over a}}\left ({k_\bot\over a}\right)\right]^2
       \delta(E - \omega)
        \eqn\this$$

    If the limit $E \rightarrow 0$ were taken at this point, we would
obtain zero emission, which would be the correct result for a static
source in the vacuum.  Because we are interested in the FDU thermal
bath corresponding to the Minkowski vacuum, we first add the induced
emission rate to the spontaneous emission rate of Eq.~\this, and then
integrate over $\omega$, to obtain the total emission rate in the
thermal bath
$$   dW^{em;therm}_{k_x,k_y} = {q^2 \over 4\pi^3 a}
   \sinh(\pi E/a) \left [{1\over e^{2\pi E/a} - 1} + 1\right ]
      \left|K_{i{E\over a}}\left ({k_\bot\over a}\right)\right|^2
     \eqno\eq $$
If we now take the limit $E \rightarrow 0$, only the contribution from
the induced emission survives, giving
$$   dW^{em;therm}_{k_x,k_y} = {q^2 \over 8\pi^3a}
      \left|K_0\left ({k_\bot\over a}\right)\right|^2
     \eqn\emit $$
The absorption rate in the thermal bath is equal to the induced
emission rate.  Because both rates are independent of the sign of
$k_x$ and $k_y$, we can simply double the above result to obtain
$$   dW^{tot;therm}_{k_x,k_y} = {q^2 \over 4\pi^3a}
      \left|K_0\left ({k_\bot\over a}\right)\right|^2
     \eqn\emit $$
which is indeed equal to the Minkowski result \minkemit\ for the
emission rate.


\appendix

	In this appendix, we study some general properties of
classical scalar
radiation. First, we consider a system of scalar charged point particles
interacting with the scalar field $\phi(x)$ which
they produce. The total action is
$$\eqalign{S &= \int d^4x\sqrt{g}{1\over 2}(\nabla\phi)^2 + \sum_n m_n
\int d\gl
\left [-g_{\mu\nu}(x_n(\gl))
{dx^\mu_n(\gl)\over d\gl}{dx^\nu_n(\gl)\over d\gl}
	\right ]^{1/2}   \cr
&\qquad + \sum_n q_n\int d\gl\ \phi(x_n(\gl))
\left [-g_{\mu\nu}(x_n(\gl))
{dx^\mu_n(\gl)\over d\gl}{dx^\nu_n(\gl)\over d\gl}
	\right ]^{1/2}   }
		\eqn\eq    $$
Varying the action with respect to metric gives the energy-momentum
tensor (in Minkowski space)
$$\eqalign{T^{\mu\nu}_{tot}(x) &= \p^\mu\phi\p^\nu\phi -
{1\over 2}\eta^{\mu\nu}
\tag     \cr
&\qquad + \sum_n\int d\tau_n\left [m_n + q_n\phi(x_n)\right ]
{dx_n^\mu\over d\tau_n}
{dx_n^\nu\over d\tau_n}\gd^{(4)}(x - x_n)     \cr
&\equiv T^{\mu\nu} + T^{\mu\nu}_{matter}    }
		\eqn\eq     $$
The conservation of $T^{\mu\nu}_{tot}$ can be explicitly verified using
the equations of motion.

	Next we study the radiation field and power of a single point
scalar charge moving along the path $\bx_s(t)$.
The Li\'enard-Wiechert potential is
\def\e{{q\over 4\pi}}
\def\E{{q^2\over 4\pi}}
\def\br{{\bf r}}
$$ \phi(\bx, t) = \e\sqrt{1 - \bv^2}\ {1\over r - \bv\cdot\br}
		\eqn\eq    $$
where $\br \equiv \bx - \bx_s(t') ,\ \bv \equiv \bv(t') ,
\ t' \equiv t - r$. 	We have
$$ {\p t'\over \p t} = {1\over 1 - \bv\cdot\hat\br} \ , \qquad
\nabla t' = -{\hat\br\over 1 - \bv\cdot\hat\br}
		\eqn\eq   $$

Introducing $ s \equiv 1 - \bv\cdot\hat\br$ and
keeping only the leading $1/r$ dependence, we obtain
the radiation fields
$$\p_t\phi(\bx,t) = {\p\phi\over \p t'}{\p t'\over \p t} =\e{1\over rs^3}
\left [ (\dot\bv\cdot\hat\br)\sqrt{1 - \bv^2} - s{\bv\cdot\dot\bv
\over \sqrt{1 - \bv^2}}\right ]
		\eqn\eq    $$
$$\nabla\phi(\bx,t) = {\p\phi\over \p t'}\nabla t'= -\e{1\over rs^3}
\left [ (\dot\bv\cdot\hat\br)\sqrt{1 - \bv^2} - s{\bv\cdot\dot\bv
\over \sqrt{1 - \bv^2}}\right ]
		\eqn\eq    $$
where all quantities on the right-hand side are to be evaluated at $t' =
t - r$. The total power radiated is
$$\eqalign{ P(t') &= \int r^2 d\Omega T^{rt}(\bx,t)
{dt\over dt'} = \int r^2d\Omega \hat\br\cdot\p^t\phi\nabla\phi{dt\over dt'}
  \cr
&= (\e)^2\int d\Omega{1\over s^5}\left [ (\dot\bv\cdot\hat\br)\sqrt{1 - \bv^2}
 - s{\bv\cdot\dot\bv\over \sqrt{1 - \bv^2}}\right ]^2   }
		\eqn\avii   $$
where the integration is over a large sphere along the forward light-cone
of the source and centered at $\bx_s(t')$.
The integration can be carried out by
introducing a coordinate system such that
$\bv\cdot\hat\br = v\cos\tg , \dot\bv\cdot\bv = v|\dot\bv|\cos\alpha ,
\dot\bv\cdot\hat\br = |\dot\bv|(\cos\tg\cos\alpha + \sin\tg\sin\alpha\cos
\varphi)    $.  One obtains
\def\gg{\gamma}
$$ P = \E{1\over 3}
{d^2x^\mu\over d\tau^2}{d^2x_\mu\over d\tau^2}
= \E{1\over 3}[\gg^4\dot\bv^2 + \gg^6(\bv\cdot\dot\bv)^2]
		\eqn\eq     $$
This is half the electromagnetic value.
The result can also be obtained by studying the non-relativistic
limit of Eq.\avii\ and then using Lorentz invariance.


\refout

\end